# Tunable-Bandwidth White Light Interferometry using Anomalous Dispersion in Atomic Vapor: Theory and Experiment

G.S. Pati, M. Salit, K. Salit, and M.S. Shahriar
*Department of EECS, Northwestern University,*
*Evanston, IL 60208*

## Abstract

Recently, the design of a white-light-cavity has been proposed using negative dispersion in an intra-cavity medium to make the cavity resonate over a large range of frequencies and still maintain a high cavity build-up. This paper presents the demonstration of this effect in a free-space cavity. The negative dispersion of the intra-cavity medium is caused by bi-frequency Raman gain in an atomic vapor cell. A significantly broad cavity response over a bandwidth greater than 20 MHz has been observed. The experimental results agree well with the theoretical model, taking into account effects of residual absorption. A key application of this device would be in enhancing the sensitivity-bandwidth product of the next generation gravitational wave detectors that make use of the so-called signal-recycling mirror.

Optical cavities with high reflectivity mirrors are now used for signal recycling in advanced gravitational wave interferometers (such as GEO 600 and the Advanced LIGO)[1,2,3]. This resonantly enhances the extremely small side-band (SB) amplitudes of gravity waves. The recycling improves the shot-noise limited sensitivity at SB frequencies due to the signal buildup inside the cavity. While the GEO 600 does not use arm-end cavities, under the Advanced LIGO program the signal recycling mirror is going to be added to a system that has arm-end cavities, yielding the same degree of improvement [4,5]. However, this improvement comes at the cost of reduced detection bandwidth.

In standard optical cavities like the ones described above, there is always a trade-off between the cavity build-up and the bandwidth. Frequency responses of the gravitational interferometers[6] containing recycling cavities have to be optimized for different types of sources such as bursts, chirps from coalescing binaries, stochastic backgrounds and continuous signals. While detecting gravity waves, one would ideally like to maximize the build-up without compromising the bandwidth, for stellar events that emit gravity waves spread over a range of frequencies (up to a few KHz). It is possible to design optical cavities with a high build-up factor but without the inverse loss of bandwidth by tailoring the path length of the cavity as a function of frequency such that the round-trip phase shift remains an integer multiple of $2\pi$. This kind of frequency-independent path length change can be achieved by using a negative dispersion intra-cavity medium with a dispersion slope corresponding to a group index value close to zero [7,8,9]. Such a cavity can simultaneously resonate over a range of frequencies without sacrificing the cavity build-up, and is called a White-Light Cavity (WLC). It has recently been shown that the use of a pair of diffraction gratings inside the cavity is not able to produce a WLC [10,11].

Proposals to construct such cavities using atomic phase coherence include strongly driven double-$\Lambda$ or double-gain systems that can provide controllable negative dispersion between gain or transparent line centers without absorption[7,8,9]. A gain based intra-cavity dispersive medium is usually preferred since the residual gain can compensate for other losses in various optical elements inside the cavity. In another development, recently a small high-Q ($> 10^9$) white-light resonator has been realized by using a dense and continuous mode spectrum of a whispering gallery mode resonator[12]. However, such a resonator is incompatible with free-space gravitational wave laser interferometers due to poor light coupling and throughput. It also is limited by the fact that it is not easily tunable after fabrication. In contrast, if the double-gain technique is employed, the cavity bandwidth can be tuned as needed by changing the gain separation and adjusting the gain parameter. Another advantage of the double-gain technique is that the gain in resonant excitations can be observed at different wavelengths in a wide variety of materials, including inorganic and semi-conducting photorefractive materials. Therefore, the technique can be extended to different operating wavelengths, for example, λ=1.064 μm in LIGO, with the proper choice of the material medium.

In this paper, we present the details of the experimental realization of a WLC using bi-frequency Raman gain in an intra-cavity atomic medium. Our experiment demonstrates the feasibility of controlling the dispersion slope such that the dispersion can exactly compensate for the frequency dependence of vacuum wavelength and produce a WLC, in good agreement with simulations. Before presenting the results, we first present a simple model summarizing the essential aspects of the WLC.

Consider an optical cavity of path length L, containing a medium of length $\ell$. The effect of medium dispersion can be seen by considering the round-trip phase shift at a frequency $\omega = \omega_o + \Delta\omega$ away from the cavity resonance frequency $\omega_o$:

$$\phi(\omega_o + \Delta\omega) = (\omega_o + \Delta\omega).(L-\ell)/c + (\omega_o + \Delta\omega).n(\omega_o + \Delta\omega).\ell/c$$
$$\simeq N(2\pi) + \left[(n_g - 1)\ell/L + 1\right].(L/c).\Delta\omega + O(\Delta\omega^2) + ..., \quad n_g = 1 + n_1.\omega_o \quad (1)$$

where $n_1$ is the first-order term in the Taylor expansion of the medium refractive index n(ω) at $\omega = \omega_o$, N is a large integer number, and $n_g$ is the group index. As can be seen from eqn.1, it is possible to choose $n_1$ such that φ(ω) becomes independent of frequency, to first-order in Δω. This happens for a value of $n_1 = -(1/\omega_o)(L/\ell)$, which corresponds to $n_g = 1 - L/\ell$, a small negative dispersion slope. In a situation where the medium completely fills the entire cavity length, $n_1 = -1/\omega_o$ and $n_g$ has a null value. This is the ideal white-light condition, under which many frequencies around $\omega_o$ will resonate simultaneously in the cavity. Thus, the cavity response gets broadened without sacrificing the cavity build-up and a WLC can be realized.

As discussed earlier, although there is no dephasing away from $\omega_o$ to first-order in Δω, dephasing due to higher-order terms causes the cavity response to drop. This eventually limits the linewidth of the WLC to a finite value. One can evaluate the WLC linewidth (γ') in comparison to the empty cavity from the following relation (see the appendix for derivation):

$$\frac{\gamma'}{\gamma} = \beta/[1 + \{(n_g - 1) + n_3 \omega_o (\gamma')^2\}.\ell/L], \quad \beta = \sin^{-1}\left[\frac{1-R\rho}{2\sqrt{R\rho}}\right]/\sin^{-1}\left[\frac{1-R}{2\sqrt{R}}\right], \quad n_3 \equiv \frac{1}{6}.\frac{\partial^3 n}{\partial \omega^3}\bigg|_{\omega=\omega_o} \quad (2)$$

where γ is the empty cavity linewidth, and ρ = exp(-αℓ), where α is the overall loss coefficient that causes additional linewidth broadening in the presence of the medium. Here, the magnitude of $|n_1.(\Delta\omega)^2.(\ell/L)|$ is considered small in comparison to the third-order dephasing term $|n_3\omega_o.(\Delta\omega)^3.(\ell/L)|$. The contribution due to the second-order coefficient $n_2$ is ignored, since the dispersion profile for an atomic resonance is anti-symmetric. The linewidth of the WLC under ideal white-light condition ($n_g = 1 - L/\ell$) can be estimated from eqn. 2 as $\gamma' = [(\beta\gamma L)/(n_3 \omega_o \ell)]^{1/3}$.

If the negative dispersion is generated using Raman gain doublets, the third-order slope ($n_3$) at the center frequency $\omega_o$ can be approximated as $n_3 = -2.n_1/\Gamma^2 = (2/\omega_o\Gamma^2).(L/\ell)$, where $\Gamma$ is the separation between the gain lines. The linewidth of WLC can then be simplified to $\gamma' = [\beta.\gamma.\Gamma^2/2]^{1/3}$. Thus, a large value of ($L/\ell$) implies that a large, negative slope is needed for the white-light condition, while the bandwidth of the WLC remains unchanged. The expression also implies that in order for the white light effect to be evident, $\Gamma$ must be much greater than $\gamma$. If we use $\gamma \sim 1$ MHz, $\Gamma = 7.95$ MHz and assume $\beta = 1$, the linewidth of the WLC is estimated to be 3.16 MHz. Of course, for applications to the LIGO system, we can design the values necessary for the desired bandwidth.

Transparent spectral gain regions and large, negative dispersion associated with gain doublets have been demonstrated in earlier experiments[13,14,15]. On the other hand, our experiment requires extremely small value of linear negative dispersion slope, typically $O(10^{-15})$ rad$^{-1}$ sec (equivalently, a small value of $n_g$) in order to satisfy the white-light condition. We have achieved this condition experimentally under low gain. In order to observe an appreciable white-light effect, the separation between the gain lines has been set larger than our actual cavity linewidth.

The experimental setup for the white-light cavity is shown in fig. 1(a). A 10 cm long Rubidium vapor cell is placed inside a 100 cm long cavity, which consists of four mirrors with two plane mirrors as input and output ports, and two concave mirrors. The empty cavity finesse and linewidth are about 100 and 3 MHz, respectively. One pair of cavity input and output ports was used to lock the cavity length at a desired resonant frequency. The probe and pump beams are obtained from a CW Ti:Sapphire laser (linewidth ~ 1 MHz). Figure 1(b) shows the $5S_{1/2}$, F=2 and F=3 ground states and $5P_{3/2}$ excited state manifolds on the D2 line in Rb$^{85}$ that constitute a $\Lambda$-type configuration for Raman excitation. An incoherent optical pump from a diode laser is used to create a population inversion between the ground states.

Two pump beams of different frequencies are injected into the cavity to generate closely spaced gain lines. These pump beams are derived using two different acousto-optic modulators (AOMs). The negative dispersion is observed in the intermediate region between the gain lines. The Raman pumps and the optical pump beams, polarized orthogonally to the probe, are all combined with the probe beam inside the cavity in the co- and counter-propagating directions, respectively, using two intra-cavity polarizing beam splitters (PBS's) before and after the cell (fig. 1a). The frequency difference between the probe and the center of the pump lines is initially matched to the ground state splitting in Rb$^{85}$ (3.0357 GHz). The gain, peaking at two different probe frequencies, is observed by scanning the probe frequency around the center frequency of the pump lines using an AOM set in a double-pass configuration. The magnetically shielded vapor cell is heated to nearly a steady temperature of 60°C using bifilarly wound coils that produce a negligible axial magnetic field.

For optimum gain, the average frequency of the probe and the pump fields is detuned below Doppler resonance where the probe does not get significantly absorbed in the cell even in the absence of the pump beams. The cavity is made resonant at the selected probe frequency $\omega_o$ by adjusting its length. The cavity length is then locked using a servo signal generated from the cavity output produced by a resonating locking beam sent through the other cavity port (fig.1). The frequency of the lock beam is set at many multiples of cavity free spectral range (FSR) away from the probe frequency $\omega_o$.

We first measured directly the dispersion as seen by the probe field under the double-gain condition, using a heterodyne technique. During this measurement, the cavity resonance was interrupted by placing a flipper mirror in the probe beam path inside the cavity. For heterodyne measurements, an auxiliary reference wave was produced by frequency shifting a fraction of the probe beam outside the cavity using a 40 MHz AOM. A low noise rf mixer and a low-pass frequency filter were used to demodulate the rf signal from the detectors that allowed us to measure accurately the dispersion in the atomic medium under the gain resonances. Very small negative dispersion slopes ($n_1 \sim 3 \times 10^{-16}$ rad$^{-1}$ sec) have been achieved with gain as low as 3 dB (or smaller). When such a low gain condition prevails inside the cavity, it also avoids other undesirable effects such as self-oscillation and bistable behavior that are normally observed in the cavity response in the presence of a stronger gain medium.

The effect of the dispersion on the cavity resonance is observed by scanning the probe frequency around the cavity resonance ($\omega_o$). Fig. 2(a) shows a sequence of cavity resonances for different frequency separations between the gain lines. Each time with an increasing frequency separation, the gain was adjusted by controlling the pump intensity to observe a white-light effect in probe transmission. The cavity linewidth in fig. 2a(i) in the absence of gain (pumps turned off) is found to be approximately 5 MHz, which is somewhat wider than the empty cavity line-width of 3 MHz. This is due to extra loss introduced by intra-cavity elements and residual medium absorption. We also see that in fig. 2a(ii) the probe transmission under white-light conditions is somewhat lower than the peak cavity transmission in the absence of gain. We attribute this reduction in peak transmission to possible additional absorption loss in probe transmission caused by optical pumping effects resulting from detuned Raman pumps. Numerical simulations, described in the next paragraph, confirm this suggestion. The simulations also show that there is a concomitant reduction in the cavity build-up factor, by about 30%. As the gain separation is increased (figs. 2a(iii, iv, v)), there is no further drop in the cavity transmission, and, therefore, no further reduction in the cavity build-up factor. Note, however, that for these three cases, the cavity transmission is not uniform over the WLC bandwidth. This is due to our limitation in optimizing the gain required to reach the white light condition ($n_g = 1 - L/\ell$). The simulations again confirm this, showing how a uniform response would have been achieved with higher gain. In future experiments, the additional loss contibuting to the overall drop in the cavity-buildup factor can be suppressed by using more efficient optical pumping. The constraint on the maximum achievable gain can also be circumvented using different experimental parameters such as the atomic density, the cell length, and the bandwidths of the AOMs used for generating the pumps.

For comparison, fig. 2(b) shows the simulated examples of WLC response for an internal cavity build-up of $O(10^2)$ and $(L/\ell) = 10$. The empty cavity resonance linewidth in fig. 2b(i) is chosen to match closely the linewidth of fig. 2a(i) observed experimentally under no-gain conditions. For the cavity response corresponding to fig. 2a(ii), we choose a gain separation of 8 MHz, and a width of ~ 2 MHz for each gain peak. The magnitude of the gain is then adjusted to yield the condition for $n_g \approx -9$. Under these conditions, a broad cavity resonance similar to the data in fig. 2a(ii) is observed. In order to obtain a closer match with the experiment, we also introduce a medium loss of coefficient 0.0005 cm$^{-1}$, which reduces the WLC build-up factor by nearly 30%. The resulting response is shown in fig. 2b(ii). For increased frequency separations, between the gain lines, while keeping the gain amplitude unchanged, non-uniform responses in the cavity transmission similar to the experimental results in fig. 2a(iii, iv, v) are observed, as shown by the solid line traces in figs. 2b(iii, iv, v). When the gain amplitude is increased in order to match the white light condition $(n_g = 1 - L/\ell)$, the non-uniformity is largely removed, as shown by the dotted line traces in figs. 2b(iii, iv, v). Fig. 2c shows a comparison of the measured linewidth for white-light response from our experimental data with the estimated linewidth from our model. They agree reasonably well within the measurement error bars. Finally, fig. 2d shows an ideal WLC response where a cavity build-up factor of 2000 is maintained.

In a separate study, we have also demonstrated experimentally the effect of positive intracavity dispersion corresponding to slow light. For this case, we have also shown how the system becomes less sensitive to perturbations[16]. Conversely, we have shown how the WLC is more sensitive to perturbations, and how this property may be used to enhance the sensitivity of absolute rotation measurement using a ring laser gyroscope[17].

The WLC described above was demonstrated using light at 780 nm, and rubidium atoms. The idea can be easily extended to Zeeman sublevels within these transitions to realize an effective three-level system, which in turn can be used to demonstrate the type of WLC described above. One example is the Zeeman sublevel based EIT we had demonstrated earlier[18]. Another example is the electromagnetically induced transparency observed in Neon gas[19]. Thus, it may easily be possible to find vapor based transitions that would yield a WLC at 1064 nm, as required for the LIGO system. Alternatively, it is possible to realize such a WLC using non-linear wave mixing in photorefractive media[20]. This work was supported in part by the Hewlett-Packard Co. through DARPA and the Air Force Office of Scientific Research under AFOSR contract no. FA9550-05-C-0017, and by AFOSR Grant Number FA9550-04-1-0189.

# Appendix

The linewidth for the WLC is derived by defining the following dephasing parameter for the cavity response

$$D_\varphi = \frac{\omega}{c_o}\left[(L-\ell) + \ell.n(\omega)\right] - \frac{\omega_o}{c}L$$

$$= \frac{L}{c_o}\left[\omega(1-\ell/L) + \ell/L.n(\omega).\omega - \omega_o\right] \quad (A1)$$

$$\approx \frac{L}{c_o}\cdot\left\{\delta\omega + \frac{\ell}{L}\left[n_1\omega_o\delta\omega + n_3\omega_o(\delta\omega)^3\right]\right\}$$

where $\omega_o$ is the cavity resonance frequency and a Taylor expansion of $n(\omega) \simeq n(\omega_o) + \left.\frac{\partial n}{\partial \omega}\right|_{\omega=\omega_o}(\omega-\omega_o) + \frac{1}{6}\left.\frac{\partial^3 n}{\partial \omega^3}\right|_{\omega=\omega_o}(\omega-\omega_o)^3 = 1 + n_1\delta\omega + n_1(\delta\omega)^3$ is used to include the effect of the medium dispersion. For bifrequency Raman gain, the second-order dispersion is considered to be absent since the dispersion profile is anti-symmetric around $\omega=\omega_o$.

Defining a similar dephasing parameter for an empty cavity, it can also be seen that for an empty cavity, $\delta\omega$ corresponds to its linewidth $\gamma$ when $D_\varphi$ equals to $\pi/F$, where F is the cavity finesse. Thus for a medium filled cavity, one can use eqn. (A1) to obtain the ratio of the linewidth for a medium-filled cavity to that for an empty cavity as

$$\frac{\pi}{F}\cdot\frac{c_o}{L} = \gamma \approx \gamma'\left\{1 + \frac{\ell}{L}\left[(n_g - 1) + n_3\omega_o(\gamma')^2\right]\right\}$$

$$\Rightarrow \quad \frac{\gamma'}{\gamma} = \frac{1}{1 + \frac{\ell}{L}\left[(n_g - 1) + n_3\omega_o(\gamma')^2\right]} \quad (A2)$$

It is important to note that in the above expression, the effect of residual loss due to absorption (considered frequency independent over cavity resonance) or scattering in the medium has not been included. This can be done as follows.

For an empty ring cavity, the cavity transmission is defined as

$$I_o(\omega) = \frac{T^2}{1 + R^2 - 2R\cos\varphi(\omega)}, \quad \varphi(\omega) = \omega.\frac{L}{c_o} \quad (A3)$$

where T (R) are the geometric mean of the intensity transmissivities (reflectivities) of the cavity input-output couplers. The effect of additional loss on the cavity linewidth[21] can be modeled by defining the cavity transmission in the presence of loss as

$$I_o(\omega) = \frac{T^2}{1 + R^2\rho^2 - 2R\rho \cdot \cos\varphi(\omega)}, \rho = e^{-\alpha\ell} \qquad (A4)$$

where $\alpha$ is the medium absorption coefficient. One can estimate the linewidth for the empty cavity by defining $\varphi_{1/2} = (\gamma/2).(L/c_o)$ for which cavity transmission $I_o$ is equal to $I_{max}/2$, where $I_{max} = \frac{T^2}{(1-R)^2}$. Using eqn. (A3), one obtains $(1-\cos\varphi_{1/2}) = \frac{(1-R)^2}{2R}$, which gives the expression for the empty cavity linewidth: $\gamma = \frac{4c_o}{L}\sin^{-1}\frac{1-R}{2\sqrt{R}}$. Note that the expression for the empty cavity linewidth $\gamma$ also agrees with the familiar expression $\gamma = 2\pi.(FSR/F)$, where FSR is the free-spectral range given by $c_o/L$ and $F = \frac{\pi\sqrt{R}}{1-R}$ when the value of R is close to unity.

Similarly, in the presence of loss as described in eqn. (A4), one can obtain the expression the broadened cavity linewidth: $\gamma" = \frac{4c_o}{L}\sin^{-1}\frac{1-R\rho}{2\sqrt{R\rho}}$. Thus, we can define a parameter, $\beta$ ($\equiv \frac{\gamma"}{\gamma} = \sin^{-1}\frac{1-R\rho}{2\sqrt{R\rho}} \Big/ \sin^{-1}\frac{1-R}{2\sqrt{R}}$), which embodies the effect of loss on the cavity linewidth[21]. Since we assume that the residual loss is independent of frequency over the effective range of non-zero dispersion, we can include the effect of this loss by simply multiplying the expression in eqn. (A2) by this parameter, $\beta$, which results in eqn. 2 in the body of the paper.

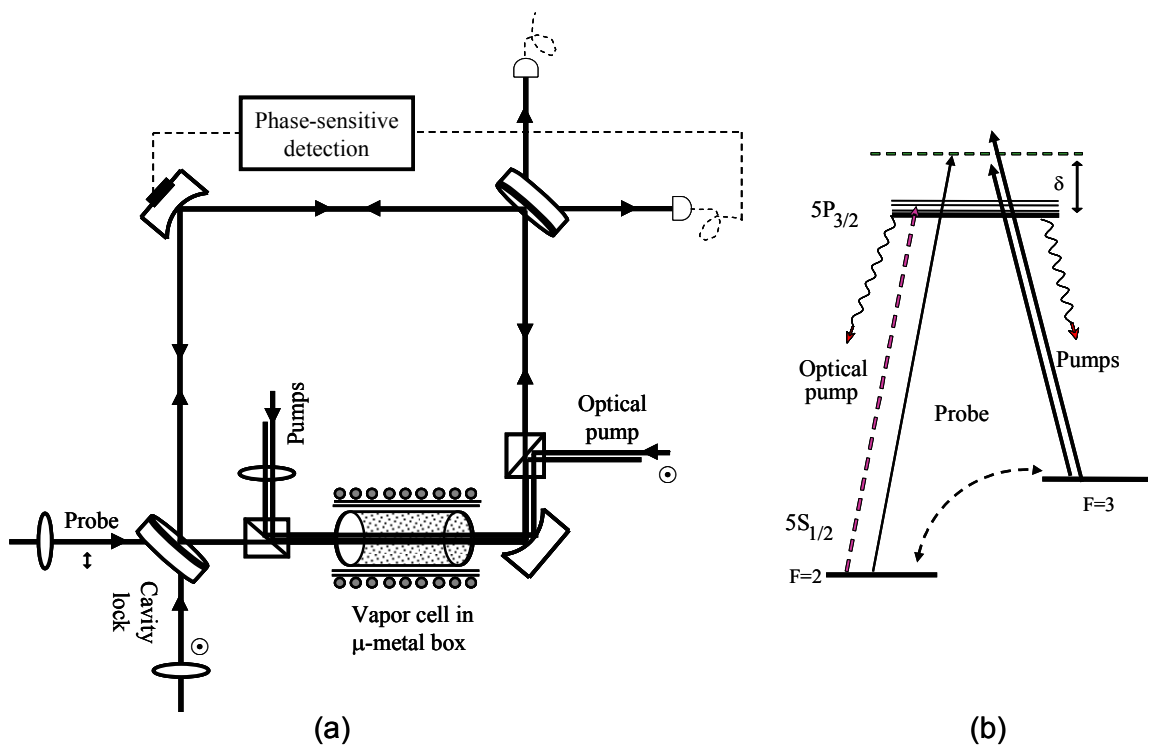

Fig.1 (a) Schematic of the experimental set-up for the white-light cavity (b) Energy diagram showing detuned Raman excitation in D2 line $^{85}$Rb to produce bi-frequency gain at probe frequencies.

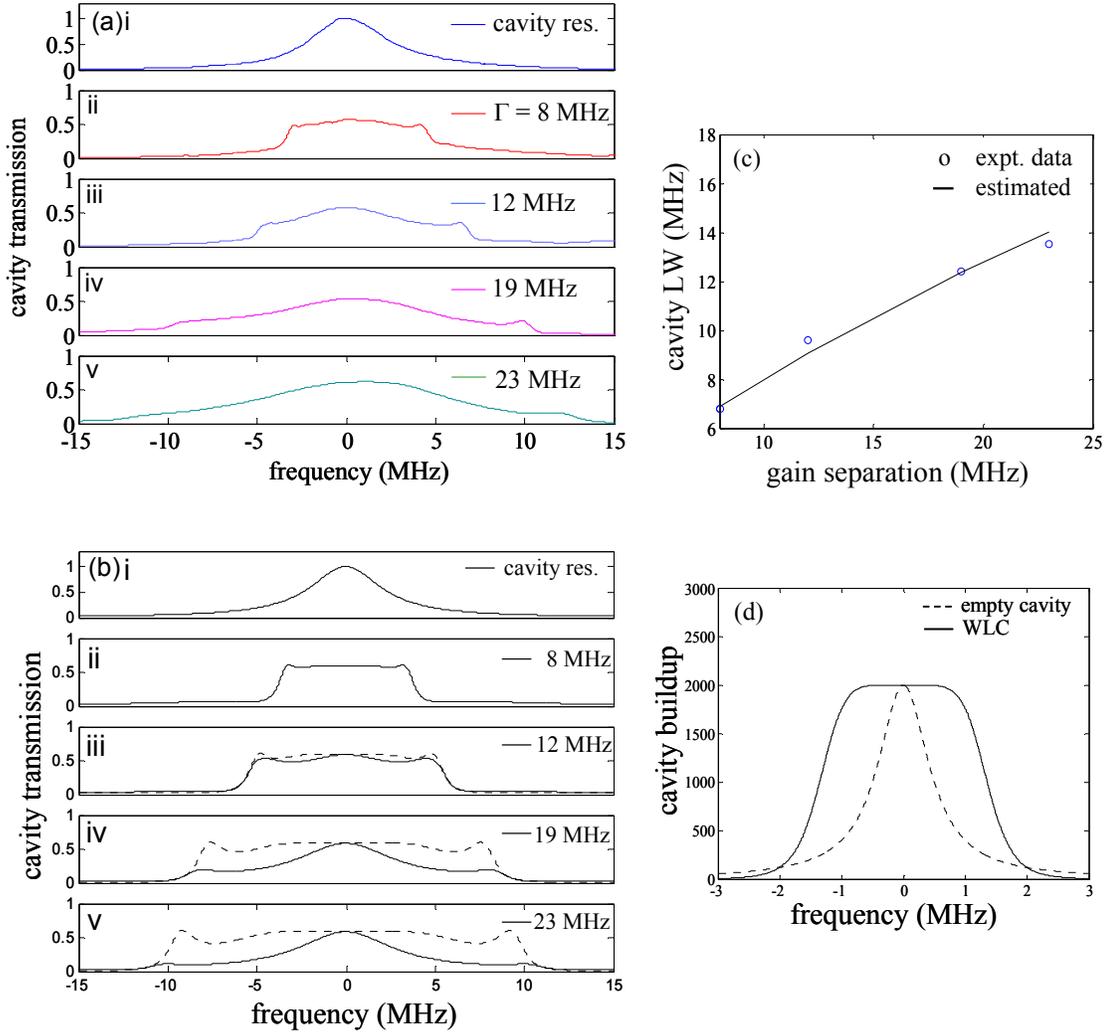

Fig. 2 (a) Experimental results showing broadened cavity response for gain doublets with varying gain separation, (b) simulated cavity resonances for different gain line separations showing close agreement with the experimental results. The dispersion slope exactly satisfies the white-light condition ($n_g = -9$) for $\Gamma = 8$ MHz. The solid lines in traces three, four and five correspond to $n_g$ values -1.95, 0.42 and 0.71 respectively. The dotted line traces in the same plots show the resonances for the exact white-light condition ($n_g = -9$) for these separations, which were achieved in simulation by increasing the gain nearly by factors 2.25, 5.65 and 8.29, respectively, and the gain linewidths by the square root of the gain factors, with respect to the gain condition at $\Gamma = 8$ MHz, (c) comparison between the estimated WLC linewidth and the actual linewidth measured from the experimental data, (d) simulated WLC response without loss, showing white-light effect at $n_g = -9$ for $\Gamma = 7.95$ MHz. The cavity buildup (x 2000) with the WLC is maintained. Note that under loss-free conditions, cavity transmission is proportional to the build-up factor.